\documentclass[twocolumn,pre,showpacs]{revtex4}
\usepackage{graphicx}
\usepackage{bm}

\begin{document}

\title{
 Hidden heat transfer in equilibrium states implies
 directed motion in nonequilibrium states
}

\author{Teruhisa S. Komatsu}%
\affiliation{Department of Physics, Gakushuin University,
 Mejiro, Tokyo 171-8588, Japan}%
\altaffiliation[Present address: ]%
{Department of Arts and Sciences, University of Tokyo,
 Komaba 3-8-1, Meguro, Tokyo, Japan 153-8902}%
\author{Naoko Nakagawa}%
\affiliation{Department of Mathematical Sciences,
 Ibaraki University, Mito 310-8512, Japan}%

\date{\today}

\begin{abstract}

We study a class of heat engines including Feynman's ratchet,
 which exhibits a directed motion of a particle in nonequilibrium steady states
 maintained by two heat baths.
We measure heat transfer from each heat bath separately, and average
 them using a careful procedure that reveals the nature of the
 heat transfer associated with directed steps of the particle.
Remarkably we find that steps are associated with nonvanishing heat transfer
 even in equilibrium, and there is a quantitative relation between this
``hidden heat transfer'' and the directed motion of the particle.
This relation is clearly understood in terms of the ``principle of heat 
 transfer enhancement'', which is expected to apply to a large class of 
 highly nonequilibrium systems.
\end{abstract}

\maketitle

\begin{figure}
\begin{center}
\includegraphics[scale=0.7]{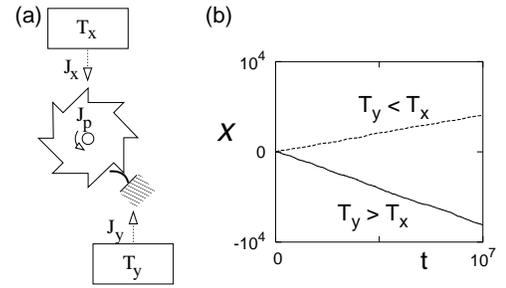}
\end{center}
\caption{
(a) Schematic figure of Feynman's ratchet composed of
 the sawtooth shaped ratchet wheel and the pawl.
The system is attached to two heat baths,
 the wheel to $T_x$ and the pawl to $T_y$.
The temperature difference between $T_x$ and $T_y$
 produces the nonvanishing steady rotation $J_p$,
 which can be used to execute external work.
The variable $x$ denotes the angle of the wheel.
There are degenerate stable positions of $x$ corresponding to
 the ratchet potential wells (there are eight in this figure).
Heat flows $J_x$ and $J_y$ are defined as positive
 when the system absorbs energy from each heat bath.
(b) Typical trajectories of $x$ in NESS.
The system exhibits directed motions,
 i.e. unidirectional rotations on average.
The solid lines and the broken lines show the data for the
 conditions $2T_x=T_y$ and $T_x=2T_y$, respectively, in Model II.
We clearly see that the direction of the motion depends on the 
 temperature difference.
}
\label{fig:ModelTraj}
\end{figure}

To understand universal features of various nonequilibrium phenomena in
 nature is still a widely open problem.
In spite of considerable interest and efforts
 (see \cite{Oono} and references therein),
 we are still very far from obtaining a universal framework
 even for nonequilibrium steady states.
We thus believe it desirable to study prototypical systems that vividly
 demonstrate properties which are essential to nonequilibrium states.
Heat engines including Feynman's ratchet \cite{Feynman}
 of Fig. \ref{fig:ModelTraj}(a), which convert
 thermal fluctuations into a directed mechanical motion,
 may be such a prototypical system.
These engines have also been investigated as Brownian motors \cite{Review-Ratchet}.
Although the Curie principle \cite{Curie} suggests that a spatial
 asymmetry of the system allows a spatially asymmetric motion of
 the particle in nonequilibrium, vivid physical pictures of the mechanism
 of the engines have been missing.
The purpose of the present letter is to develop such a universal
 physical picture.

As is well-known, Feynman's ratchet consists of a wheel and a pawl
 attached to separate heat baths, and exhibits a mechanical rotation
 \cite{Feynman}.
Although Feynman introduced this model to illustrate the impossibility
 of designing a heat engine with the efficiency exceeding the Carnot limit,
 it has been shown that Feynman's ratchet cannot attain the Carnot efficiency
 \cite{Sekimoto,Parrondo}.
In the present study, we focus on a more primitive and hopefully fundamental
 aspect of the ratchet problem, namely, the direction of the rotation.
Feynman designed the ratchet so that the wheel is likely to rotate
 in one direction.
But, as Feynman himself pointed out, the wheel can rotate in the opposite direction
 depending on the temperatures of the baths (see Fig. \ref{fig:ModelTraj}(b)).
This fact indicates that the direction of the motion is a delicate issue
 which requires a careful consideration.

In the present work, we shall reveal that there is a precise statistical
 quantity that determines the direction of the motion.
Such an investigation has a practical importance in higher dimensional
 ratchet-like problems, where the preferred direction of the motion
 is far from manifest.
We shall point out a deep relation between the directed motion
 and ``hidden heat transfer'' that takes place
 in the {\em equilibrium state}.
Our main conclusion is summarized in the ``principle of heat 
 transfer enhancement''.
We believe that our findings cover a large class of heat engines,
 and shed new light on universal features of nonequilibrium steady
 states.

\paragraph*{Models:}
Feynman's ratchet may be realized as discrete stochastic models
 \cite{Jarzynski} or as continuous models described by
 a set of Langevin equations \cite{Sekimoto}.
We here concentrate on the latter type.
The model consists of one translational degree of freedom $x$
 corresponding to the angle of the wheel
 (in the following we call it the position of a particle)
 and other degrees of freedom $\bm{y}$ describing the
 mechanical interaction between the pawl and the ratchet wheel.
The degrees $x$ and $\bm{y}$ are in contact with
 heat baths with temperature $T_x$ and $T_y$, respectively.
By suitably choosing $T_x$ and $T_y$, we can study
 the behavior of the model in equilibrium or
 in nonequilibrium steady states (NESS).
The time evolution of the system is described
 by the set of Langevin equations,
\begin{eqnarray}
m_x \ddot x &=&
 -\gamma_x \dot x +\sqrt{2\gamma_x T_x}\,\xi_x(t)
-{\partial U(x,\mbox{\boldmath$y$})}/{\partial x},\nonumber\\
m_{\bm{y}} \ddot{\bm{y}} &=&
 -\gamma_{\bm{y}} \dot{\bm{y}}
 +\sqrt{2\gamma_{\bm{y}} T_y}\,\bm{\xi_y}(t)
-{\partial U(x,\bm{y})}/{\partial{\bm{y}}}
 \nonumber,\\
\label{eq:model}
\end{eqnarray}
 where the $x$-asymmetric interaction potential $U(x,\bm{y})$
  has a translational invariance in $x$ with period $l=1$
\footnote{
To be precise, the translational invariance means 
$U(x+l,\mathrm{R}_l\bm{y})=U(x,\bm{y})$.
Here $\mathrm{R}_l$ is the operator which translates
  $\bm{y}$ by $l$ along the wheel.
In case of Model I, $\mathrm{R}_l y=y$.
In case of Model II, $(\mathrm{R}_l \bm{y})_i=y_{i-1}+l$.
}.
Because the potential has stable fixed points in every period,
 equivalent binding states are aligned periodically in $x$.
$\gamma_x$ and $\gamma_{\bm{y}}$ are the friction coefficients
 (here we choose $\gamma_x=\gamma_{\bm{y}}=1$)
 and $\xi_{*}(t)$ represent Gaussian white noises
 with a variance of unity.
In this letter, we use two concrete models to demonstrate
 the results numerically, but the results do not depend
 on the specific models.

``Model I'' is Sekimoto's version of Feynman's ratchet \cite{Sekimoto}
 which is the simplest case having only two degrees of freedom ($x$ and $y$).
The potential is given by $U(x,y)=\exp(-y+\phi(x))+y^2/2$, where 
 $\phi(x)$ is a sawtooth shaped periodic function of $x$
 \footnote{
Here the smooth function $\phi(x)=-\sin(2\pi x)/2-\sin(4\pi x)/12+1/2$
 is used for convenience of numerical calculations.
}.
The inertia terms are neglected ($m_x=m_y=0$).

``Model II'' has higher degrees of freedom, and was introduced
 as a toy model for molecular motors
 \cite{Nakagawa_Kaneko4,Nakagawa_Komatsu,Nakagawa_Komatsu_PhysicaA}.
The particle $x$ and chain sites $\bm{y}=\{y_i\}$
 are located on a one-dimensional circle.
The interaction potential
 $U(x,\bm{y})=\sum_i [v(x-y_i)+u_1(y_i-y_{i+1})+u_2(y_i-il)]$
 is composed of three parts, namely, as
 asymmetric nonlinear potential $v$ between the particle and the chain site,
 harmonic potential $u_1$ between the neighboring chain sites
 and harmonic on-site potential $u_2$
 \footnote{
The magnitude of $u_1$ and $u_2$
 determines the stiffness of the system,
 which is a control parameter used for exploring model space.
Please refer to \cite{Nakagawa_Kaneko4,Nakagawa_Komatsu}
 for the explicit form of the potential functions and
 to \cite{Nakagawa_Komatsu_PhysicaA} for the method
 to determine the basin boundary.
}.
We set $m_x=m_{\bm{y}}=1$.

\paragraph*{Hidden heat transfer in equilibrium:}
In shorter time scales, the particle (the angle of the wheel)
 is mostly bound to one of the binding states of the potential well,
 and from time to time exhibits sudden jumps (steps) to neighboring binding states
 due to thermal activation.
The probabilities of rightward and leftward steps
 must be identical in equilibrium, but are in general different in NESS.
This unbalance generates a directed motion of the particle
 in longer time scales.

The thermally activated step is an elementary process of the system.
During a single step
 (i.e., a jump of the particle from a binding state to a neighboring state),
 the system first absorbs some amount of energy (heat)
 from the heat baths and returns it afterward.
A close investigation comparing the heat transfer associated
 with leftward steps and that with rightward ones will reveal
 the hidden heat transfer in equilibrium.

In order to examine the heat transfers from respective heat baths,
 let us define the time-dependent heat flows
 $\hat{J}_x(t)$ and $\hat{J}_y(t)$ (energy absorbed into
 the system per unit time) for each trajectory
 by using the stochastic energetics method \cite{Sekimoto,Sekimoto_Takagi_Hondou} as
\begin{eqnarray}
\hat{J}_x(t) =
 [ - \gamma_x \dot x +\sqrt{2\gamma_x T_x}\,\xi_x(t) ]
 \circ \dot x,
\nonumber\\
\hat{J}_y(t) =
 [ - \gamma_{\bm{y}} \dot{\bm{y}}
 +\sqrt{2\gamma_{\bm{y}} T_y}\,\bm{\xi_y}(t) ]
 \circ \dot{\bm{y}},
\label{eqn:heatflow}
\end{eqnarray}
 which should be time integrated with the Stratonovich interpretation.

The heat flows $\hat{J}_x(t)$ and $\hat{J}_y(t)$
 exhibit large fluctuation and hardly allow any physical interpretation
 as they are.
In order to detect the heat transfers
 associated with the thermally activated step,
 we introduce a carefully {\it conditioned ensemble average} as follows.
For each step, we shift the time variable so that at time $t=0$
 the particle moves from the basin of a binding state
 to the basin of a neighboring state.
Then we perform ensemble averaging of 
 $\hat{J}_x(t)$ and $\hat{J}_y(t)$,
 separately for rightward steps and leftward steps.
The averaged quantities are denoted as 
 $J_x^\mathrm{R}(t)$, $J_x^\mathrm{L}(t)$,
 $J_y^\mathrm{R}(t)$, and $J_y^\mathrm{L}(t)$,
 where $J_x^\mathrm{R}(t)$ is the heat flow
 from $T_x$ associated with rightward steps, and so on.

\begin{figure}
\begin{center}
\includegraphics[scale=0.7]{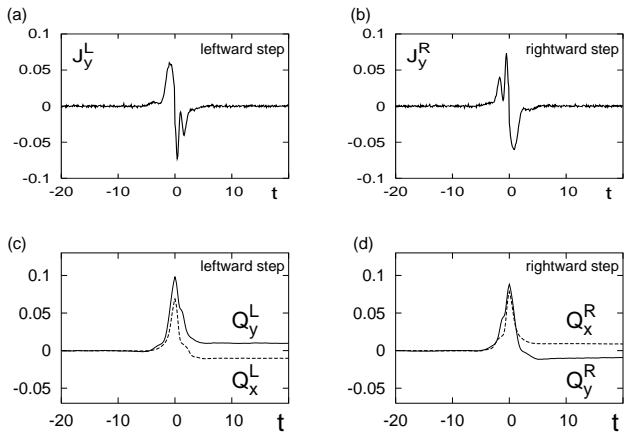}
\end{center}
\caption{
Time sequences of the averaged heat flow $J_y(t)$ over conditioned ensembles
($3.8\times 10^5$ samples) for (a) leftward step and (b) rightward step
 in the equilibrium ($T_x=T_y=0.05$) for Model II,
 where the energy barrier height is about $3.1 T_x$.
For each step, time is shifted
 so that the particle crosses basin boundaries at $t=0$.
Heat transfers $Q_*(t)$ for (c) leftward
 and (d) rightward steps respectively, which is obtained by the
 integration of $J_*(t)$ from $t_0=-20$ (see Eq.(3)).
}
\label{fig:JQeq}
\end{figure}

Fig.\ref{fig:JQeq}(a)(b) shows these averaged heat flow profiles
 in {\it equilibrium state}.
The system absorbs energy from the heat baths before the step ($t<0$), 
 and dissipates it after the step ($t>0$).
The heat flow rapidly converges to zero
sufficiently after or before the step,
 and net heat flow appears only around the step.

The heat transfer, i.e., time integrated heat flow,
 is defined as
\begin{equation}
Q_x^\mathrm{R}(t)
\equiv\int_{t_0}^{t}{\rm d}t' J_x^\mathrm{R}(t'),
\label{eq:qdef}
\end{equation}
 where $t_0<0$ is chosen so that
 $J_x^\mathrm{R}(t)$ is negligible for $t\le t_0$.
The quantities
$Q_x^\mathrm{L}(t)$, $Q_y^\mathrm{R}(t)$,
 and $Q_y^\mathrm{L}(t)$ are defined similarly.
Fig.\ref{fig:JQeq}(c)(d) shows the heat transfer in equilibrium state.
We here notice the significant fact that energy absorbed from one heat bath
 before the step is not returned to the same heat bath after the step.
This implies that, {\it even in equilibrium}, each step carries heat
 from one heat bath to the other.
Although the existence of such heat transfer may look surprising, 
 it does not contradict with any of the thermodynamics laws
 \footnote{
Comparing the profiles for rightward and leftward steps,
 we can find that the equalities
$J_y^\mathrm{R}(t)=-J_y^\mathrm{L}(-t),
J_x^\mathrm{R}(t)=-J_x^\mathrm{L}(-t)$
 hold. 
We also find that rightward and leftward steps occur with exactly the same
 probability in equilibrium.
Indeed these properties are necessary consequences of the reversibility
 of equilibrium states.
That we have confirmed these properties can be regarded as
 a sign of reliability of our numerical calculations.
}.
The nonvanishing heat transfer is observed only when we treat rightward
 and leftward steps separately.
The word ``hidden heat transfer'' denotes such heat transfer.
Of course, when averaged over all steps in both the directions,
 there is no heat transfer in equilibrium.

In the following, we denote by $q^\mathrm{eq}$ the hidden heat transfer
 from $T_x$ to $T_y$ in equilibrium
 associated with rightward steps. Note that we have
\begin{equation}
q^\mathrm{eq} \equiv Q_x^\mathrm{R}(\infty) = -Q_x^\mathrm{L}(\infty)
=-Q_y^\mathrm{R}(\infty) = Q_y^\mathrm{L}(\infty)
\end{equation}
as is seen in Fig.\ref{fig:JQeq}(c)(d).
The nonvanishing hidden heat transfer $q^\mathrm{eq}$ in equilibrium
 comes from the asymmetry of the interaction potential $U(x,\bm{y})$
 and the resulting dynamics.
Since such an asymmetry is not a special feature
 of the present models but a rather generic one,
 we expect that similar nonvanishing hidden heat transfer
 in equilibrium states is found in a wider class of systems.

\paragraph*{Heat transfer and directed motion in NESS near equilibrium:}
The ``hidden heat transfer in equilibrium'' explored in the above
 determines the direction of the particle motion in NESS
 as we shall now discuss.
Let us define the response coefficient $\chi_\mathrm{p}$
 of the particle flow to the temperature difference as
\begin{equation}
\chi_\mathrm{p} = \lim_{\Delta\beta \rightarrow 0}
 J_\mathrm{p}(\beta-\Delta\beta/2,\beta+\Delta\beta/2) / \Delta\beta,
\label{eq:chip}
\end{equation}
 where
 $J_\mathrm{p}(\beta_x,\beta_y)
   =\langle\dot{x}_\mathrm{p}\rangle$ 
 is the particle current for
 $T_x=1/\beta_x$ and $T_y=1/\beta_y$
 (where $\Delta\beta=1/T_y-1/T_x$ and $2\beta=1/T_y+1/T_x$).
One of course has $ J_\mathrm{p}(\beta,\beta)=0$.
From the sign of $\chi_\mathrm{p}$,
 the direction of the particle flow is specified.
For example, when $\chi_\mathrm{p}>0$
 right-oriented flow appears for $\Delta\beta>0$,
 i.e., $T_x>T_y$.

\begin{figure}
\begin{center}
\includegraphics[scale=0.8]{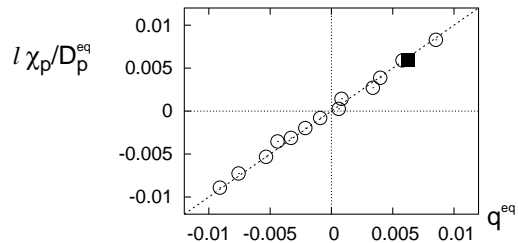}
\end{center}
\caption{
Hidden heat transfer $q^\mathrm{eq}$ at equilibrium
 and 
 $l \chi_\mathrm{p} /D_\mathrm{p}^{\rm eq}$.
This reveals that there is a nontrivial relation
(\protect\ref{eq:qchip}) between the motion of the particle in a NESS
and ``hidden heat transfer'' in equilibrium.
Results for the two models, Model I (filled square) at $\beta=3$
 and Model II (open circles) at $\beta=20$, are shown.
For the latter model, the stiffness parameter of the model
 is varied to explore model space broadly so that we have
 many points for the same temperature conditions.
}
\label{fig:qchipKc}
\end{figure}

As shown in Fig.\ref{fig:qchipKc}, we found
 a remarkable relation between $\chi_\mathrm{p}$ and
 the hidden heat transfer $q^\mathrm{eq}$,
\begin{equation}
q^\mathrm{eq} = \frac{l\,\chi_\mathrm{p}}{D_\mathrm{p}^\mathrm{eq}}
\label{eq:qchip}
\end{equation}
 where $D_\mathrm{p}^\mathrm{eq}$ is the diffusion constant of the particle
 in equilibrium states and $l$ is the period
 of the interaction potential in $x$.
The relation (\ref{eq:qchip}) does not only relate
 the directions of particle motion
 and ``hidden heat transfer'',
 but shows that these two are related in a quantitative manner.
In the systems of heat engines, the directed particle flow
 is induced by a ``field'' associated with temperature differences.
Based on linear response theory \cite{Kubo},
 the quantity $\chi_\mathrm{p}$, which is the particle flow
 divided by a temperature gradient, is expressed in terms of
 a time correlation function  between the particle flow $\dot{x}$
 and the heat flow $(\hat{J}_x-\hat{J}_y)/2$ at equilibrium.
Evaluating this expression under the assumption that each step
 occurs independently, we can derive Eq. (\ref{eq:qchip})
 \cite{Komatsu_Nakagawa_prep}.

Eq.(\ref{eq:qchip}) implies that
 a step in the direction of the particle flow
 enhances heat transfer from the hotter to the colder heat bath.
For example,
 considering the case $\chi_\mathrm{p}>0$ and $T_x>T_y$,
 the particle flows toward right from the definition (\ref{eq:chip}).
At the same time, Eq.(\ref{eq:qchip}) implies $q^\mathrm{eq}>0$, which
 means that rightward steps carry heat $q^\mathrm{eq}$
 from $T_x$ to $T_y$, i.e., from hotter to colder,
 while leftward steps carry the same amount of heat in the opposite direction.
We conclude that the direction of the particle motion is chosen so that
 to enhance the heat transfer between the two baths.
This ``principle of heat transfer enhancement'' may look quite natural
 and reasonable.

\paragraph*{Heat transfer and directed motion in NESS far from equilibrium:}

Let us proceed to the observation of heat transfer far from equilibrium.
Because there is net steady heat transfer,
 we consider the {\it excess} heat transfer
 associated with steps in NESS defined as
\begin{equation}
Q_{x,\mathrm{ex}}^\mathrm{R}(t)
\equiv\int_{t_0}^{t}{\rm d}t [ J_x^\mathrm{R}(t) - \bar{J_x} ].
\end{equation}
The quantities
$Q_{x,\mathrm{ex}}^\mathrm{L}(t)$,
$Q_{y,\mathrm{ex}}^\mathrm{R}(t)$
 and $Q_{y,\mathrm{ex}}^\mathrm{L}(t)$ are defined similarly.
Here
 $J_x^\mathrm{R}(t)$ is the conditioned ensemble average of the heat flow
 and the contributions from the steady heat flow,
 $\bar{J_x}$ and $\bar{J_y}$
 satisfying $\bar{J_y}=-\bar{J_x}$,
 are subtracted, where ``bar'' means the long time average.
Note that we have
$\lim_{t\to\pm\infty} J_x^\mathrm{R}(t) = \bar{J_x}$,
 etc.

\begin{figure}
\begin{center}
\includegraphics[scale=0.7]{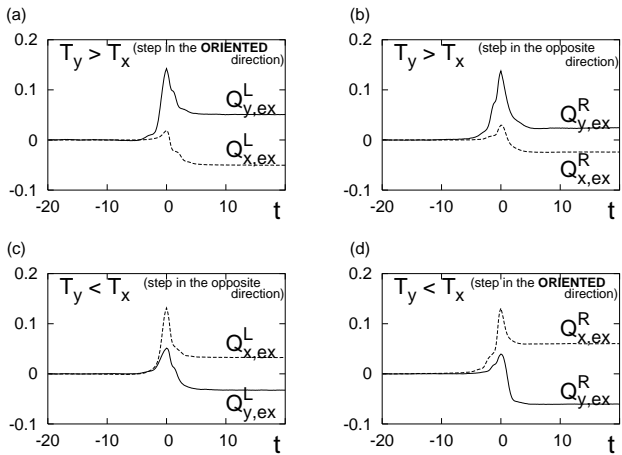}
\end{center}
\caption{
Profile of excess heat transfer in NESS of Model II.
In the upper figures with $T_y=2T_x$, particle flow leftward.
In the lower figures with $2T_y=T_x$, particle flow rightward.
}
\label{fig:Qnessflex}
\end{figure}

Fig. \ref{fig:Qnessflex} shows the excess heat transfers in NESS.
Note that, in the figure, the excess heat transfers
 associated with rightward and leftward steps have the
 same direction, namely, from the hotter bath to the colder.
However, the amounts of heat transfer are different for the rightward
 and leftward steps, and we again find that the direction which
 enhances the net heat transfer is selected.
We conclude that the ``principle of heat transfer enhancement'' holds
 also in systems far from equilibrium.

\paragraph*{Discussion:}
We have investigated a class of heat engines including Feynman's ratchet
 and found a clear relation between the directed motion in NESS
 and hidden heat transfer in the equilibrium state.
The relation (\ref{eq:qchip}),
 which universally holds for the present class, can be derived
 from the cross correlation between the fluctuation of the heat flow
 and that of the particle flow.
By focusing on thermally activated steps, we were able to characterize
 this correlation in terms of the hidden heat transfer $q^\mathrm{eq}$.
The notion of the hidden heat transfer provides us with a vivid picture
 for the elementary process of the heat engine, namely, 
 each directional step carries a heat quanta depending on its direction.
Heat transfers separately measured for each heat bath would be useful quantities
 for the investigation of other systems with multiple heat baths.
It is also interesting to explore other models of heat engine
 suitable for theoretical treatments \cite{Jarzynski,Broeck}.
The fact that the principle of heat transfer enhancement holds even
 in NESS far from equilibrium suggests that
 there can be a universal characterization of NESS.
This remains as a future problem.
We believe that our findings offer a novel universal view point
 for studying NESS,
 and will lead to a better understanding of nonequilibrium physics in general.

We are grateful to 
H. Tasaki for a critical reading of this manuscript.

\end{document}